\documentclass[10pt, conference, compsocconf]{IEEEtran}

\usepackage{hyperref}

\ifCLASSINFOpdf
  \usepackage[pdftex]{graphicx}
  \graphicspath{{../pdf/}{../jpeg/}{../png/}{./}}
  \DeclareGraphicsExtensions{.pdf,.jpeg,.png}
\else
\fi
\usepackage{fancyhdr}
\usepackage{verbatim}
\usepackage{amsmath}
\usepackage{tikz}
\usetikzlibrary{trees}

\begin{document}
%
\title{Development of a Big Data Framework for Connectomic Research}


\author{\IEEEauthorblockN{Terrence Adams}
\IEEEauthorblockA{U.S. Government\\
9161 Sterling Dr.\\
Laurel, MD 20723\\
tma2131@columbia.edu}
}


%


\maketitle

\begin{abstract}
This paper outlines research and development of a new 
Hadoop-based architecture for distributed processing 
and analysis of electron microscopy of brains. 
We show development of a new C++ library 
for implementation of 3D image analysis techniques, 
and deployment in a distributed map/reduce framework. 
We demonstrate our new framework on a subset 
of the Kasthuri11 dataset from the Open Connectome Project. 
\end{abstract}

\begin{IEEEkeywords}
brain; electron microscopy; connectome; hadoop; map/reduce; service;

\end{IEEEkeywords}

%
\IEEEpeerreviewmaketitle

\section{Introduction}

In recent years, there has been rapid growth in research on 
the human brain, as well as brains of other animals. 
The improvement and wide applicability of non-invasive techniques, 
such as fMRI and new high-resolution electron 
microscopy, are fueling new research on brain physiology, 
and brain function. 
In the past, advanced research on brain physiology was constrained 
to domain experts from the medical fields, neuroscience and 
other researchers with access to limited datasets. 
The advent of ``big data'' ecosystems, coupled 
with new brain research initiatives, 
large datasets and techniques are becoming more widely available. 
This is opening a new set of problems to a wider audience 
of professionals from computer science, mathematics and 
engineering. 

This paper outlines a framework for developing 
and deploying image-based analytics on high-resolution 
electron microscopy of animal brains. 
Our focus is design and development of a ``full-stack'' 
prototype for deploying analytics in a Hadoop map/reduce 
ecosystem. By ``full-stack'', we include interfaces and services 
that implement a pipeline from data download 
to distributed analytic processing and results output. 
This task was challenging in a short time-frame, 
so much more work is required to produce a mature system 
that can be used broadly.  This work was completed 
as part of a graduate course at Columbia University, 
called ``Big Data Analytics''.  
We plan to continue this research in the follow-on course, 
``Advanced Big Data Analytics''.



\section{Related Works}
Big Data Analytics is a broad area of research. 
By its nature, it encompasses research on large frameworks 
for processing massive quantities of data. Often the data is 
disparate and multi-modal. Finally, the goal is state-of-the-art 
application of cutting-edge analytics in a large framework. 
In recent years, there has been tremendous growth 
in development of breakthrough advanced mathematical and 
computer science methods for tackling big data. 

In computer vision, areas such as deep learning are advancing 
the state-of-the-art in significant ways. Also, for years, 
research and development of probabilistic graphical 
models, belief propagation and graph analytics have 
pushed state-of-the-art forward. 
In particular, in areas of face detection and recognition, 
probabilistic graphical models have been known to produce 
state-of-the-art accuracy, when coupled with robust 
feature extraction, and careful training. 

Currently, several centers are researching and developing 
new technologies that address brain understanding problems. 
New insight is being gained by collection massive quantities 
of high resolution imagery, and advanced analytic processing 
in large frameworks.  For this project, we were fortunate 
to leverage expertise from the Open Connectome Project. 
See \cite{OCP-web}, \cite{OCP-1} and \cite{OCP-2} for 
further details on this project.  

There is a proliferation of big data frameworks, distributed 
processing and storage architectures, as well as stream 
processing environments. Prior to undertaking analytic 
research and development, we surveyed several 
publicly-available distributed processing frameworks.  We wanted to make 
early decisions on which programming languages and libraries 
to focus on.  Below we provide a list of some of the frameworks 
that were considered for this project. We do not provide 
details on each framework here. 

\subsection{Some Publicly-Available Processing Frameworks}
Here is a list of publicly available distributed processing frameworks that were considered for this final project:  
\begin{enumerate}
\item Hadoop Map/Reduce 
\item Apache Spark 
\item Apache (Twitter) Storm 
\item Yahoo! S4 
\item IBM InfoSphere Streams 
\item MIT CSAIL Streamit 
\item ZeroC ICE 
\item FastFlow (with ZeroMQ) 
\item Muesli 
\item SkeTo 
\item RayPlatform 
\item CAF (C++ Actor Framework)
\item FlowVR 
\item OpenMP \& MPI 
\item Intel TBB 
\item Apache (Facebook) Thrift 
\end{enumerate}

\section{System Overview}

\begin{figure}[!h]
\centering
\href{http://cloud.ganita.org/images/BEDdiagram.png}{\includegraphics[width=2.5in]{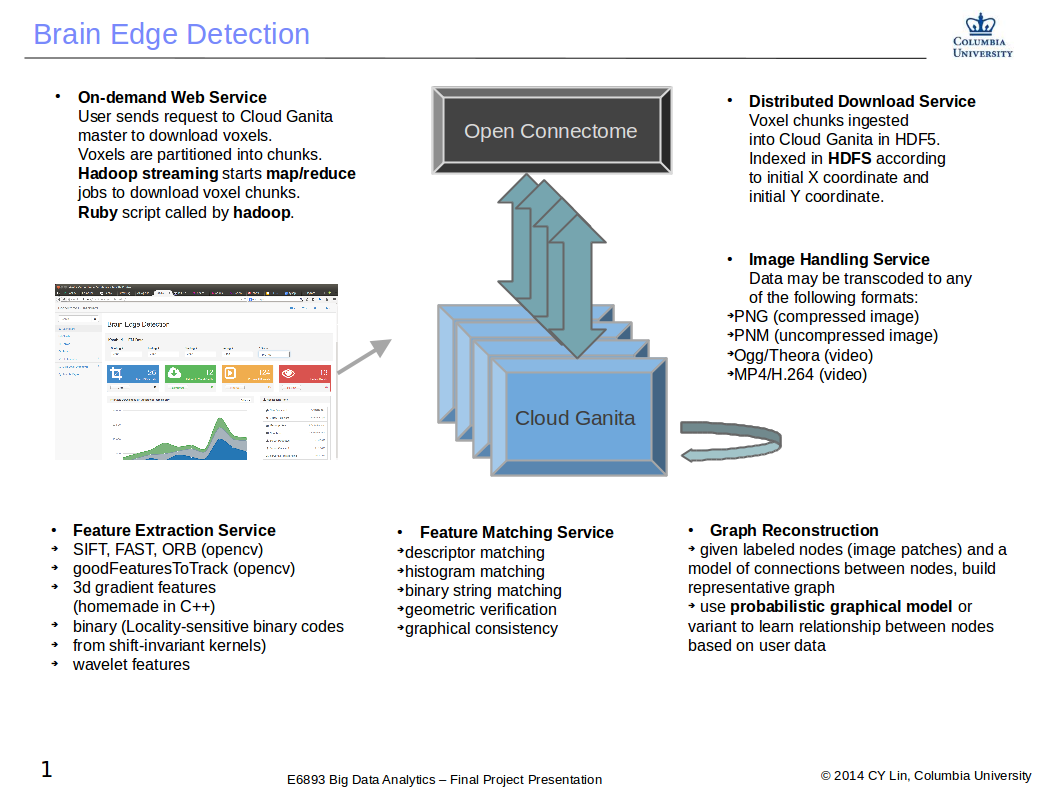}}
\caption{System Diagram}
\label{sys_diag}
\end{figure}

Developing all the components from Figure \ref{sys_diag} 
in a short time frame is very challenging. We divided 
research into the following categories:
\begin{enumerate}
\item Distributed Download Service 
\item Image Handling Service 
\item On-demand Web Service
\item Feature Extraction Service
\item Feature Matching Service 
\item Graph Reconstruction
\end{enumerate}
One or more components have been developed in each 
of the six categories above. At this time, a complete 
pipeline is not fully constructed. 
We stood-up a cluster of commodity computers 
for the purpose of this course. We continue 
to use this cluster for development, testing 
and giving demos of our services. 

Our system is called cloud ganita, 
and the following section provides more detail. 
Following that, we describe data used for this project 
and then outline various components developed 
for the course project: Brain Edge Detection. 

\subsection{Cloud Ganita}

All installations, maintenance and system administration 
were performed on a small cluster, 
stood-up for this project. 
The table below shows the role of individual nodes 
in the Cloud Ganita cluster. 

\vskip .1in
\begin{center}
   \resizebox{3.2in}{!} {
    \begin{tabular}{ | l | l | l |}
    \hline
    Computer name & Hadoop/YARN Type(s) & HBASE/Zookeeper Type \\ \hline \hline 
    ganita8 & NN, DN, NM & HM, QP, HRS \\ \hline
    ganita7 & DN, NM & RS, QP \\ \hline
    ganita6 & DN, NM & HBM, RS, QP \\ \hline
    ganita5 & RM, DN, NM, JHS & No ZK \\ \hline
    ganita4 & DN, NM, JHS & No ZK \\ \hline
    \end{tabular}
   }
\end{center}
{\scriptsize 
NN=NameNode, RM=ResourceManager,
DN=DataNode, NM=NodeManager, 
JHS=JobHistoryServer, ZK=ZooKeeper, 
HM=HBase Master, HBM=HBase Backup Master, 
RS=HBase RegionServer, QP=ZK QuorumPeer, 
HRS=HBase RESTServer }

\begin{figure}[!ht]
\centering
\href{http://cloud.ganita.org}{\includegraphics[width=2.5in]{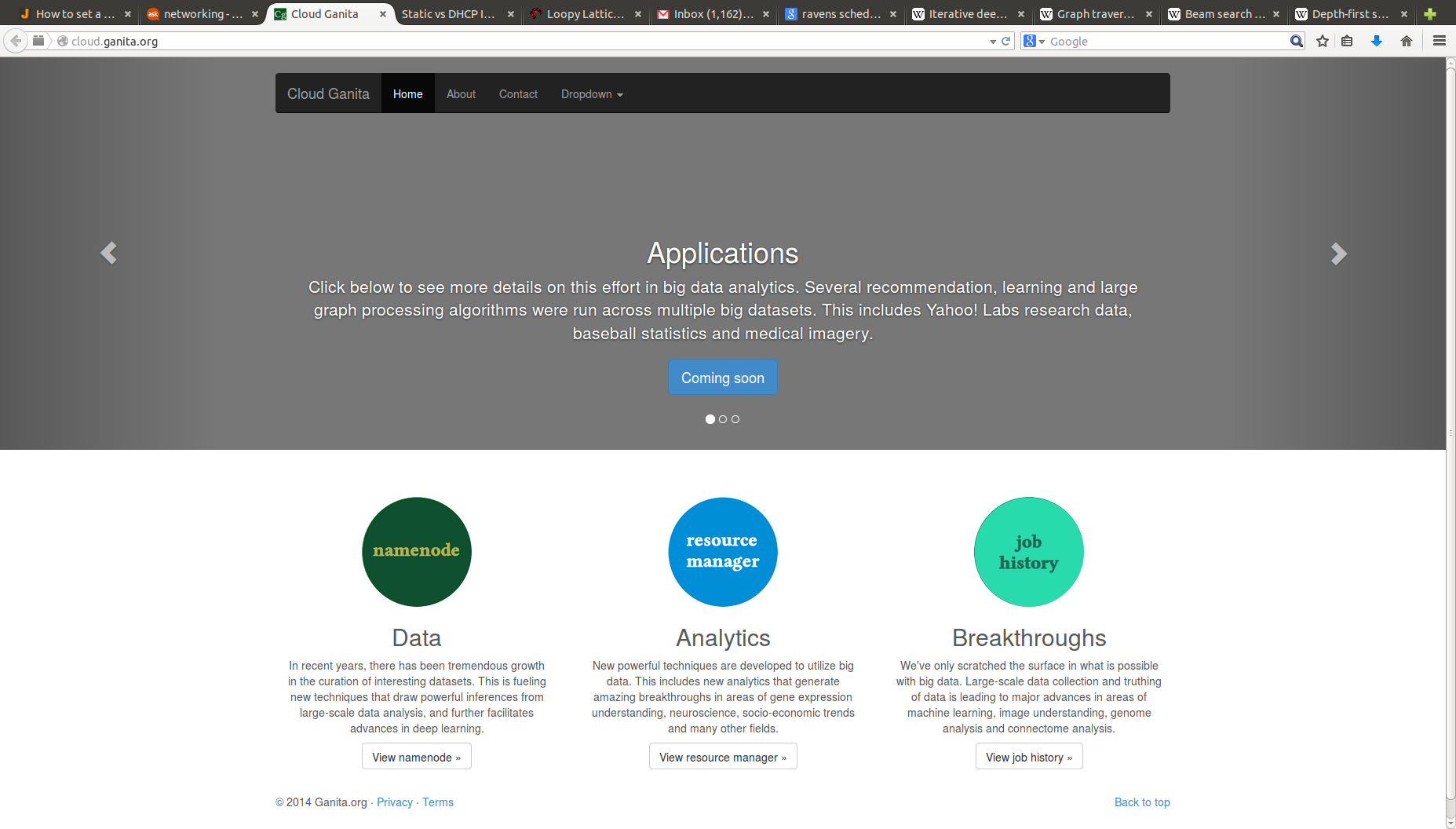}}
\caption{Cloud Ganita Web Interface}
\label{cloud-ganita1}
\end{figure}

Figure \ref{cloud-ganita1} shows a web interface 
created for this project. 
From the web interface, it is possible to view 
namenode, resource manager and job history 
on the Hadoop cluster. 
It is currently password protected.  Log in - 
username: {\bf ganita} and password: {\bf cloud}. 

\vskip .1in 

\subsection{OCP Electron Microscopy}
All of the data used for this project was 
made available by the Open Connectome Project (OCP).
\footnote{See http://www.openconnectomeproject.org/.} 
We focused on high-resolution electron microscopy 
collected by Bobby Kasthuri (BU \& Harvard). 
It is known as the Kasthuri11 dataset. 
Here is a description of this data: 
\begin{itemize}
\item OCP data is available via a RESTful interface
\item 3x3x30 cubic nanometer spatial resolution
\item 660 GB of images
\item Up to 1,000,000,000,000 pixels 
\item Focused on 10752 x 13312 x 1850 resolution 
\item $\sim$ 265 billion pixels
\item Info $\rightarrow$ http://openconnecto.me/ocp/ca/kasthuri11/info/
\end{itemize}

\subsection{System Components}
The distributed download service has a UI and allows 
a user to carve out HDF5 voxels for download and indexing 
into HDFS. 
The UI is built using html5, javascript and php. 
The php commands partition the EM blocks into 
a manageable number of slices and load 
a configuration file into HDFS. 
Then, a php command launches a Hadoop map/reduce 
job that feeds lines from the configuration 
file to a ruby script. The ruby script uses curl 
on each configuration line to start the download 
from any one of the datanodes in our cluster. 

The image handling service allows conversion 
of hdf5 image data into any of the following formats:
\begin{itemize}
\item Uncompressed imagery (PNM/PPM/PGM) 
\item PNG compressed images 
\item JPEG compressed images 
\item OGG/Theora compressed video 
\item MP4/H.264 compressed video 
\end{itemize} 
We used ffmpeg along with the plug-in codecs: 
libogg, libtheora, libx264, and the standard codecs. 

The on-demand web service is the user interface 
for automatically downloading voxels from OCP. 

The feature extraction service includes two main 
components. The first is a newly built C/C++ library 
for extracting gradient features from EM data. 
The second is integration of OpenCV into the cloud 
environment and application of standard OpenCV 
feature extractors to the converted imagery. 
OpenCV is the prevailing open package 
for deploying and running computer vision algorithms 
on image or video. 

The feature matching service is still in development. 
It will include feature matching from OpenCV, along 
with newly created matchers in a C/C++ library. 

We may also continue to explore feature extraction 
and matching in Matlab.  It enables quick development 
of new advanced algorithms. Also, OCP has built a new 
powerful API (called Cajal3d) that will work directly 
with the OCP RESTful interface. 

For graph reconstruction, we have begun development 
of a new graph processing engine in C/C++. 
This engine will be tailored to the EM brain data. 
Plans are to integrate this engine with other existing 
graphical capabilities such as System G, libDAI, etc.  

\section{Algorithms} 
Several tools and algorithms have been researched and developed 
for the purpose of processing electron microscopy. 
Only recently, have massive volumes of high-res EM scans 
of brains been available. Typically, in the medical imaging 
fields, focus has been on lower resolution 2D imagery. 

Recently, the Open Connectome Project has built 
a big data framework (using Sun Grid Engine) 
for deploying computer vision 
algorithms in a distributed manner. Also, created 
by OCP is a Matlab API for rapidly developing and testing 
new algorithms in this framework. 

We found that advanced and robust computer vision 
algorithms are developed in Matlab, C/C++, or in CUDA. 
We have made an early decision to focus on three 
main languages (or development environments) for creating 
new algorithms.  With our eye on future research 
to include development of deep learning algorithms, 
probabilistic graphical models and dynamical systems 
approaches, we will focus on a few different 
deploy-and-test environments. 
\begin{enumerate}
\item Matlab $\rightarrow$ OCP Sun Grid Engine 
\item C/C++ $\rightarrow$ Hadoop streaming \& InfoSphere Streams 
\item CUDA $\rightarrow$ GPU \& caffe 
\end{enumerate}
For this paper, we focus on research and development 
of C/C++ algorithms and deployment of some of these 
algorithms in a Hadoop map/reduce framework. 

Here is a list of the main algorithms developed for this effort:
\begin{enumerate}
\item Hadoop streaming HDF5 download service 
\item Image handling service (data conversion)
\item Feature building (C++)
\item Graph analytic starter engine (C++) 
\item Web interface for project and basic service (html5/JS)
\end{enumerate}
Also, we started testing of basic computer vision 
feature extractors from OpenCV on the EM data. 
However, since these algorithms are designed 
for 2-dimensional data (images or motion frames), 
we have started development of a new C/C++ 
library for integration of 3-dimensional 
volumetric processing of EM data. 
Initially, we focused on research and development 
of 3D gradient feature extractors. 
We developed a 3D Sobel operator, along 
with modified versions for tuning on high-res EM volumes. 

\subsection{3D Sobel Operator}
We developed a standard 3D Sobel operator for application 
on EM voxels. We used the following kernel: 
\[ K(-1) = \left( \begin{array}{ccc}
-1 & 0 & 1 \\
-3 & 0 & 3 \\
-1 & 0 & 1 \end{array} \right)\] 

\[ K(0) = \left( \begin{array}{ccc}
-3 & 0 & 3 \\
-6 & 0 & 6 \\
-3 & 0 & 3 \end{array} \right)\] 

\[ K(1) = \left( \begin{array}{ccc}
-1 & 0 & 1 \\
-3 & 0 & 3 \\
-1 & 0 & 1 \end{array} \right)\] 

Below are two images: the first shows an original EM slice, 
and the second shows the output of the 3D Sobel operator. 
\begin{figure}[!ht]
\centering
\includegraphics[width=2.5in]{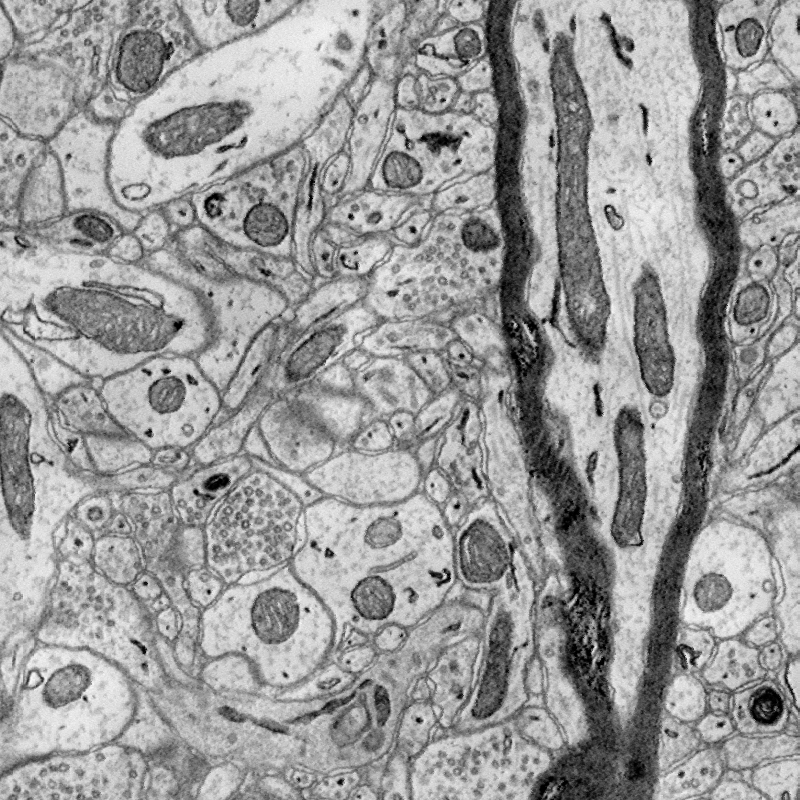}
\includegraphics[width=2.5in]{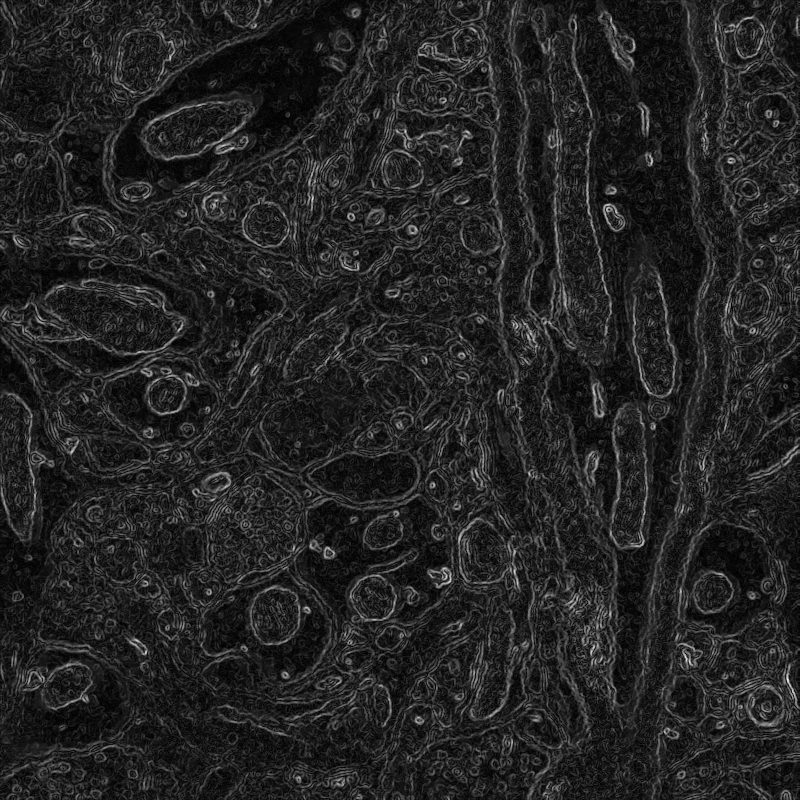}
\caption{3D Sobel Operator - Slice View}
\label{sobel3D-1}
\end{figure}
We also added the norm as a parameter and tested 
a modified Sobel operator with $L_2$ replaced 
by $L_p$ for various $p$ values. Finally, we added 
the ability to binarize the output of the 3D Sobel 
operator using mean and standard deviation-based 
thresholds. 
\begin{figure}[!ht]
\centering
\includegraphics[width=2.5in]{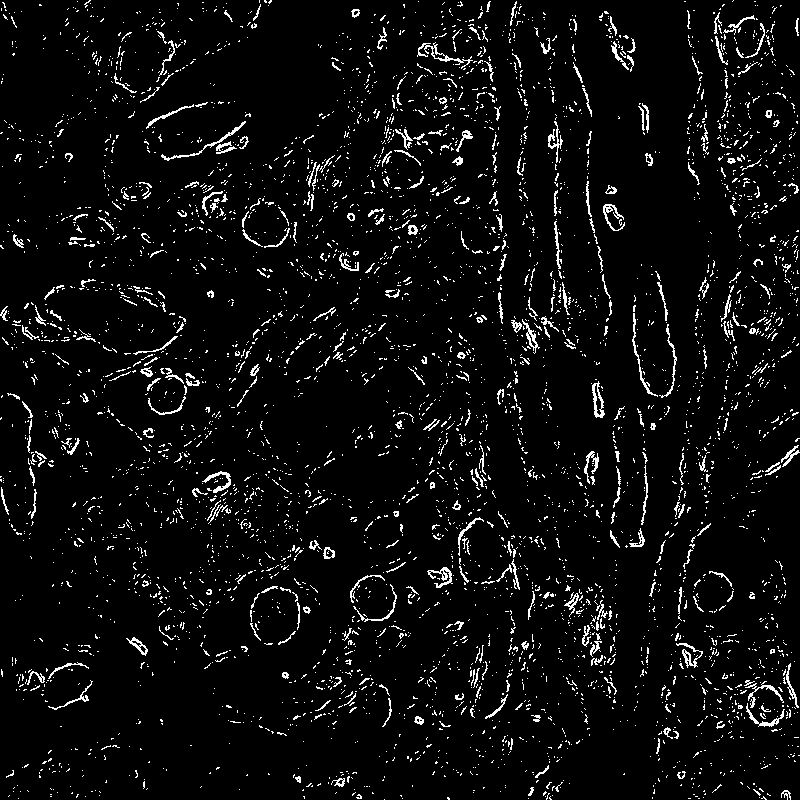}
\caption{Binarization of 3D Sobel Operator Output}
\label{sobel3D-2}
\end{figure}

Notice clear remnants of closed curves, many 
with circular eccentricity. 
We are working on an algorithm to integrate around points 
on the closed curves and output a single stable center 
point.  Our main goal is to detect and extract feature 
patches for registering regions of the connectome. 
We develop and energy function that can be optimized 
that outputs many of these center points. 
From these, a sparse graph can be constructed, and then 
further dense modeling of EM blocks and connectome regions. 
We will extract a feature vector based on cell regions, 
neurons, axons and synapses. We first build a graphical model 
based on statistically stable regions learned 
from experiments on large volumes of EM data. 

The Ganita feature extraction module consists 
of two main C++ classes: Ganita3D and GanitaImage. 
GanitaImage handles reading, writing of images, videos 
and interfacing with external image and video libraries. 
Ganita3D implements tailored 3D feature extractors such 
as the 3D Sobel filter. 

\subsection{Graph analytic starter engine}
We have developed a basic graph processing 
engine in C++. Our goal is to use this engine 
to further analyze EM data and integrate with existing 
probabilistic graphical frameworks. 
Currently, our plans are to integrate 
with libDAI which is a free and open source 
C++ library for Discrete Approximate Inference 
in graphical models. 

To store a graph, several data structures are created. 
The goal is to avoid allocating 
$(\#nodes) \times (\#nodes)$ locations in memory, 
since this may be very large.  
We create an array that stores the index of each vertex, and 
an array that stores the offset into the edge array for each vertex. 
Then all edges are packed into a single array of longs in numeric 
(vertex order, adjacency vertex order). 
Total number of memory allocations equals:
\[
\#edges + 2\times \#vertices . 
\]

With each node and edge, a property can be 
stored. For image processing, the property may 
contain features from a region of the EM block data.
As an example, the graph library can 
compute the dot product between 
two nodes (as the sum of common adjacent vertices). 
This can be computed combinatorically efficiently by scanning 
through the adjacent vertices in unison, while incrementing. 

\verbatiminput{./combinatoric_dot.cpp}

Note, a standard method for computing the dot product would 
be to store the edges in an adjacency matrix and compute 
the dot product of two binary vectors.  We do not wish 
to allocate this much memory. 

The Ganita Graph software is composed of the following 
C++ classes: 
\begin{align*}
&GanitaTrace.cpp \\ 
&GanitaNode.cpp \\ 
&GanitaGraph.cpp \\ 
&GanitaFormatter.cpp \\ 
&GanitaEdge.cpp . 
\end{align*}

\section{Software Package Description}
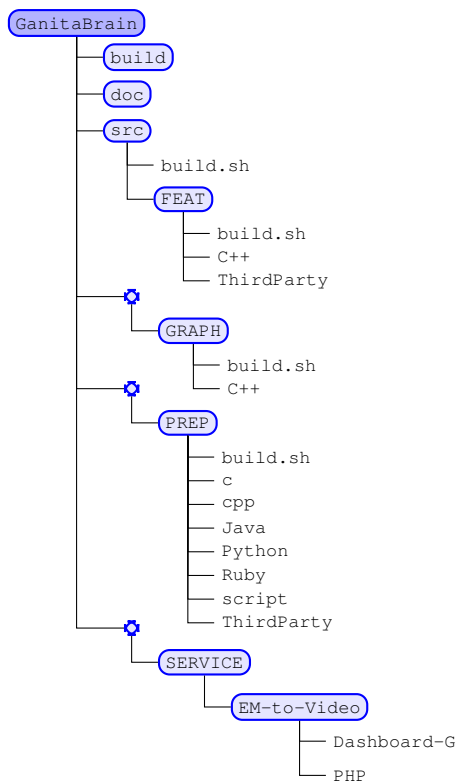
\begin{figure}[!ht]
\label{software-fig}
\tikzstyle{every node}=[thick,anchor=west, rounded corners, font={\scriptsize\ttfamily}, inner sep=2.5pt]
\tikzstyle{selected}=[draw=blue,fill=blue!10]
\tikzstyle{root}=[selected, fill=blue!30]

\begin{tikzpicture}[%
    scale=.7,
    grow via three points={one child at (0.5,-0.65) and
    two children at (0.5,-0.65) and (0.5,-1.1)},
    edge from parent path={(\tikzparentnode.south) |- (\tikzchildnode.west)}]
  \node [root] {GanitaBrain}
    child { node [selected] {build} }
    child { node at (0,-.25) [selected] {doc} }
    child { node [selected] at (0,-.5) {src}
      child { node {build.sh}}
      child { node [selected] at (0,-.2) {FEAT}
        child { node {build.sh}}
        child { node {C++}}
        child { node {ThirdParty}}
      }
    }       
    child { node at (.4,-3.2) [selected] {}
      child { node [selected] {GRAPH}
        child { node {build.sh}}
        child { node {C++}}
      }
    }
    child { node at (.4,-4.5) [selected] {}
      child { node [selected] {PREP}
        child { node {build.sh}}
        child { node {c}}
        child { node {cpp}}
        child { node {Java}}
        child { node {Python}}
        child { node {Ruby}}
        child { node {script}}
        child { node {ThirdParty}}
      }
    }
    child { node at (.4,-8.6) [selected] {}
      child { node [selected] {SERVICE}
        child { node [selected] at (0,-.2) {EM-to-Video}
          child { node {Dashboard-G}}
          child { node at (0,-.2) {PHP}}
        }
      }
    };
\end{tikzpicture}
\caption{GanitaBrain Directory Structure}
\end{figure}

A new project was created in github that contains 
some of the source code used for this project. 
The project name under github is {\bf GanitaBrain}. 
The software is structured as shown 
in Figure \ref{software-fig}. 

The software in PREP is used to prepare the EM data 
for analytic processing. This includes distributed download 
from the Open Connectome Project and conversion to various 
formats. 

The directory FEAT contains the gradient feature extractor, 
including a 3-dimensional modified Sobel operator. 
Also, included in the same class is a binarization routine. 

The root directory contains three subdirectories: 
build, doc and src. 
To build the software, a user must first install 
all dependencies.  There is a actually a long list 
of dependencies, but the most important ones are: 
libcurl, h5dump, hdf5 libraries and executables, 
ffmpeg, opencv. 
The C++ software in FEAT and GRAPH can be built without 
installing most dependencies. 
To build executables and jar files, a user should 
change directory to the root directory and enter the 
command:
\[
sh\ \ build/build.sh 
\]

\vskip .1in 
\noindent
The following figure shows the download of GanitaBrain 
from github and the successful build process. 
\begin{figure}[!h]
\centering
\includegraphics[width=2.5in]{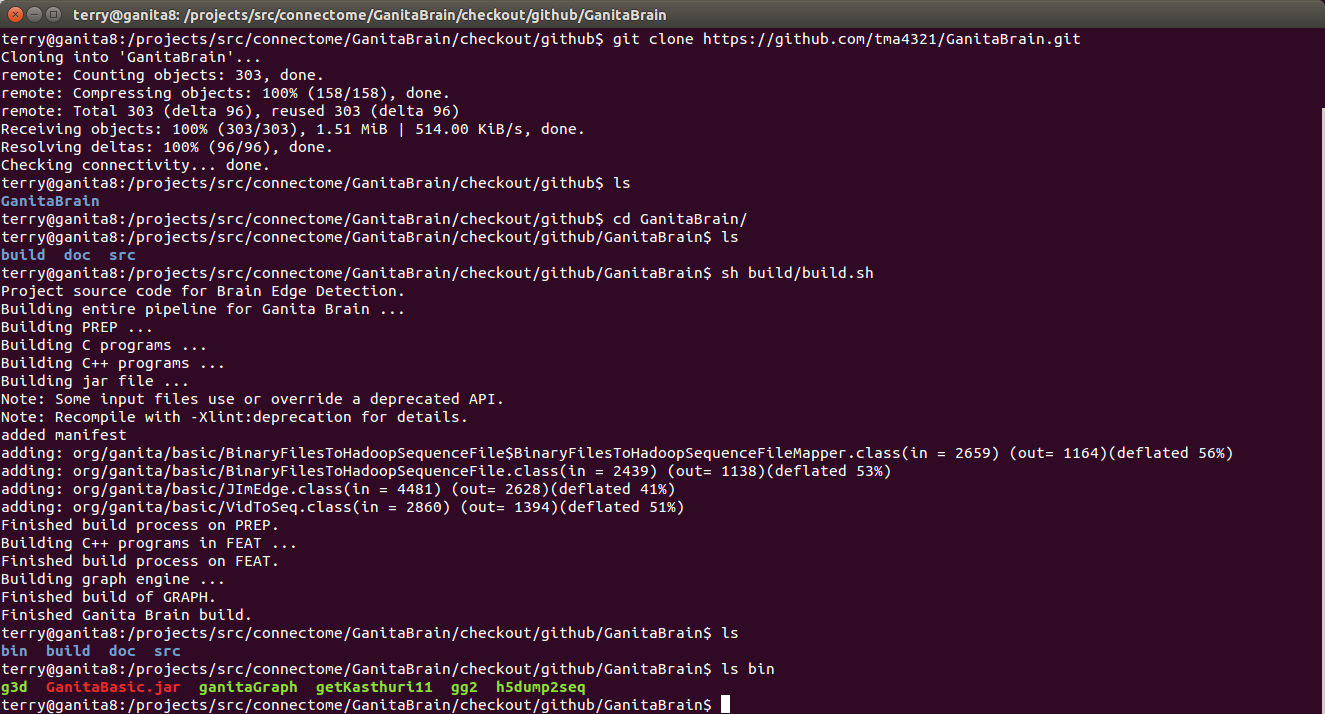}
\caption{Successful Software Build}
\label{build-1}
\end{figure}

\section{Experiment Results}

Several experiments were run using the software developed 
for this effort. 
Below is a screenshot from Cloud Ganita jobs 
that perform a distributed download, ingest and indexing 
of HDF5 imagery. 
\begin{figure}[!h]
\centering
\includegraphics[width=2.5in]{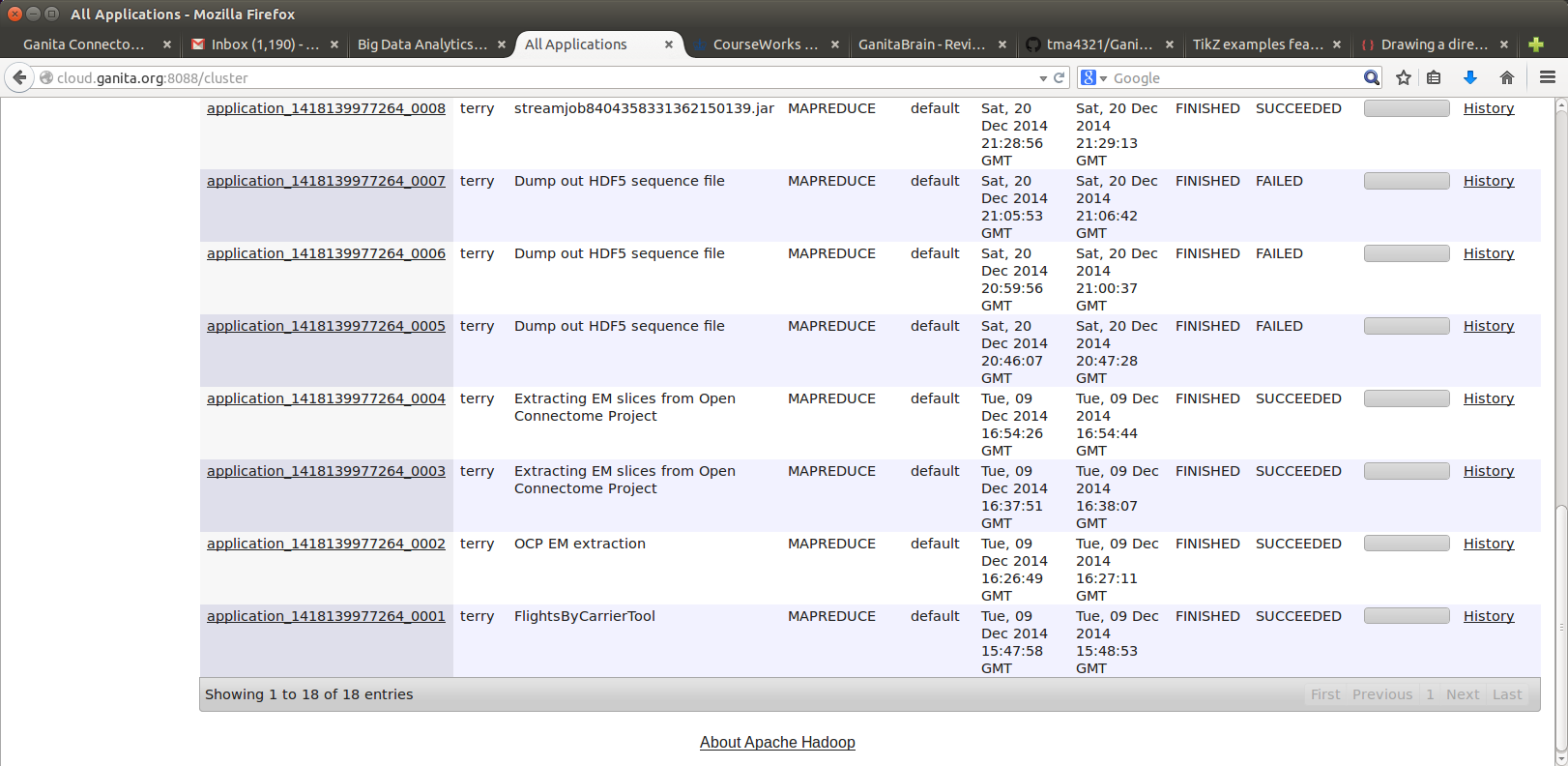}
\caption{Extraction of HDF5 from OCP}
\label{OCP-extract-1}
\end{figure}

\vskip .1in

\noindent 
Below are screenshots of Cloud Ganita namenode and datanodes. 
Following that is a screenshot of the HDFS directory 
showing EM data stored in 3 different formats: 
HDF5, SequenceFile and uncompressed PGM text files. 
\begin{figure}[!h]
\centering
\includegraphics[width=2.5in]{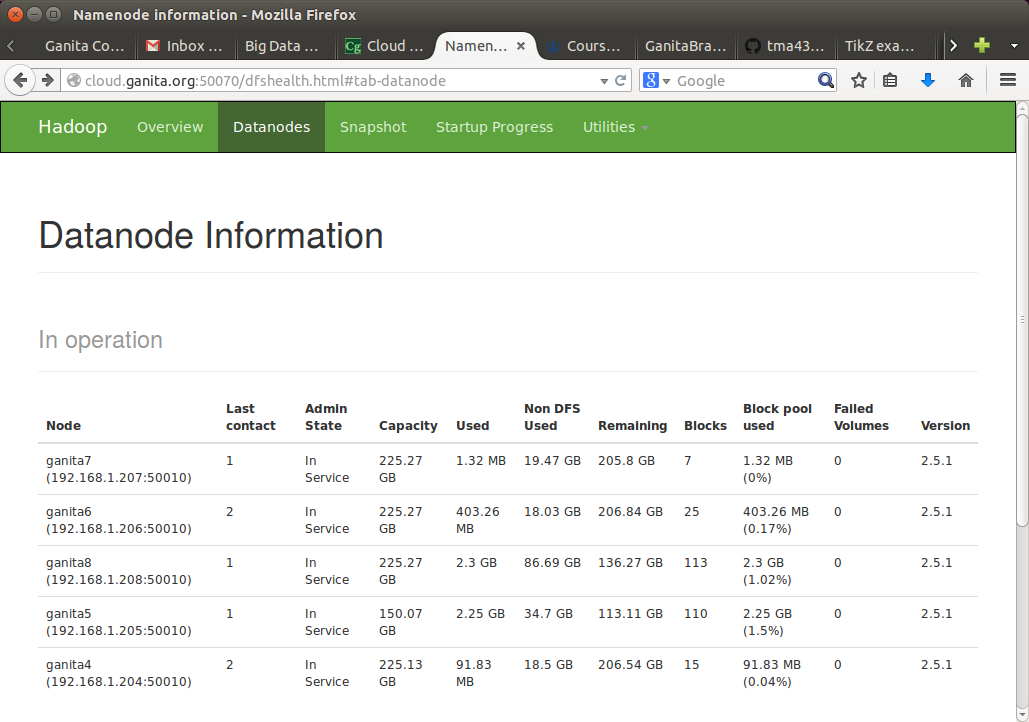}
\caption{Cloud Ganita Cluster}
\label{cluster-1}
\end{figure}

\vskip .1in

\begin{figure}[!h]
\centering
\includegraphics[width=2.5in]{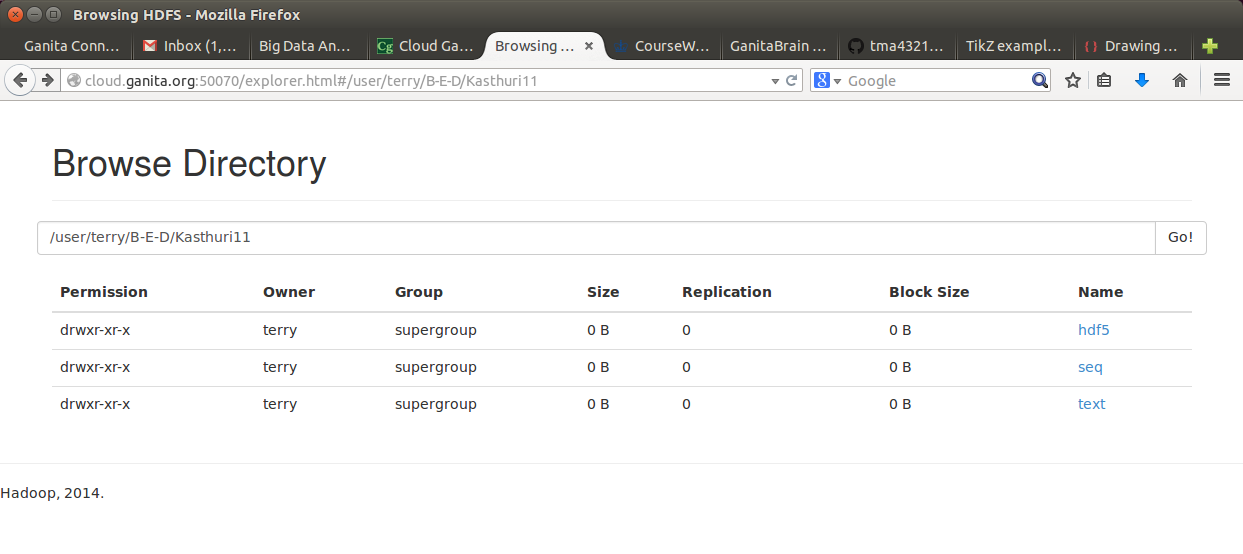}
\caption{Cloud Ganita EM Data}
\label{cloud-data}
\end{figure}
At this time, we have various analytics running 
in Hadoop/HDFS.  We have not yet collected 
ground-truthed data to test results from 
our query-by-example framework. 
This paper focuses on research to identify 
components for the big data framework, 
and development to deploy and integrate 
many of these components. 
Future research will focus on establishing 
a pipeline for processing data in an on-demand 
fashion and comparing results to ground-truth. 
An interesting task is identifying synapses 
in the Kasthuri11 dataset. 

\section{Conclusion}
For future directions, this author plans 
to focus on application of 3D feature extraction, 
labeling and training of classification using 
deep learning framework (i.e. caffe), and 
application of probabilistic graphical models. 
Focus will be on environments well suited 
C/C++ algorithm development.  Further utilization 
of Matlab will take place. 

\section*{Acknowledgment}

The author acknowledges assistance from several members 
of the Open Connectome Project, including Will Gray Roncal, 
Jacob Vogelstein and Mark Chevillet (Johns Hopkins University). 
The author did this research while a student 
in Columbia University's Big Data Analytics 
course.  The author thanks teacher, 
Ching-Yung Lin, for several insightful comments 
that will guide future research. 
\newpage



%

\end{document}